\documentclass{emulateapj}

\usepackage{graphics}
\usepackage{epsfig}
\usepackage{natbib}
\shorttitle{Far-UV Continuum in Disks I: Accretion \& H$_{2}$}
\shortauthors{France, Yang, \& Linsky}

\begin{document}


\title{The Far-Ultraviolet ``Continuum'' in Protoplanetary Disk Systems~I: Electron-Impact H$_{2}$ and Accretion Shocks\altaffilmark{*}}


\author{Kevin France\altaffilmark{1}} 
\affil{Center for Astrophysics and Space Astronomy, University of Colorado, 389 UCB,  
Boulder, CO 80309}
\author{Hao Yang} 
\affil{JILA, University of Colorado, 440 UCB, Boulder, CO 80309}

\and

\author{Jeffrey L. Linsky}
\affil{JILA, University of Colorado and NIST, 440 UCB, Boulder, CO 80309}


\altaffiltext{*}{Based on observations made with the NASA/ESA $Hubble$~$Space$~$Telescope$, obtained from the data archive at the Space Telescope Science Institute. STScI is operated by the Association of Universities for Research in Astronomy, Inc. under NASA contract NAS 5-26555.}

\altaffiltext{1}{kevin.france@colorado.edu}




\begin{abstract}

We present deep spectroscopic observations of the classical T Tauri stars DF Tau and V4046 Sgr in order to better characterize two important sources of far-ultraviolet continuum emission in protoplanetary disks.  These new {\it Hubble Space Telescope}-Cosmic Origins Spectrograph observations reveal a combination of line and continuum emission from collisionally excited H$_{2}$ and emission from accretion shocks. H$_{2}$ is the dominant emission in the 1400~$\lesssim$~$\lambda$~$\lesssim$~1650~\AA\ band spectrum of V4046 Sgr, while an accretion continuum contributes strongly across the far-ultraviolet spectrum of DF Tau.  We compare the spectrum of V4046 Sgr to models of electron-impact induced H$_{2}$ emission to constrain the physical properties of the emitting region, after making corrections for attenuation within the disk.   
We find reasonable agreement with the broad spectral characteristics of the H$_{2}$ model, implying $N(H_{2})$~$\sim$~10$^{18}$ cm$^{-2}$, $T(H_{2})$~=~3000$^{+1000}_{-500}$ K, and a characteristic electron energy in the range of $\sim$ 50~--~100 eV.
We propose that self-absorption and hydrocarbons provide the dominant attenuation for H$_{2}$ line photons originating within the disk.  
For both DF Tau and V4046 Sgr, we find that a linear fit to the far-UV data can reproduce near-UV/optical accretion spectra.  We discuss outstanding issues concerning how these processes operate in protostellar/protoplanetary disks, including the effective temperature and absolute strength of the radiation field in low-mass protoplanetary environments.  We find that the 912~--~2000~\AA\ continuum in low-mass systems has an effective temperature of $\sim$~10$^{4}$~K with fluxes 10$^{5-7}$ times the interstellar level at 1 AU.

\end{abstract}



\clearpage


\section{Introduction}

Classical T Tauri stars (CTTSs) are characterized by broad H$\alpha$ emission lines and ultraviolet (UV) spectra dominated by 
atomic and molecular features that are attributed to gas-rich disks~\citep{furlan06}.    The ages of CTTS disks (0.1~--~12~Myr; 
Isella et al. 2009; Kastner et al. 2008) indicate that the gaseous processes contributing to their observational characteristics are contemporaneous with and likely intimately connected with giant planet formation, which is thought to be mostly completed on similar timescales~\citep{alibert05}.\nocite{isella09,kastner08}  Terrestrial planet formation is thought to occur on timescales of 10~--~100 Myr~\citep{kenyon06}, when the majority of the primordial gas is in the disk has been dissipated.
Observations of the gas and dust in CTTS disks therefore probe the physical and chemical state of gas giant forming protoplanetary
systems.  

Multi-wavelength spectroscopy is a powerful tool for making quantitative measurements of the dust and gas in these systems.  
Dust in protoplanetary disks is seen most clearly through the mid- and far-infrared (IR) excess flux produced by warm grains~\citep{furlan06}, which overwhelms the narrow molecular lines when observed at low-resolution~\citep{najita10}.
Emission from hot gas is produced by magnetic activity in the atmospheres of the central stars and the accretion shocks near the 
stellar surface~\citep{krull00,ardila02}, while the protoplanetary material itself can be probed through molecular observations of these 
systems.  Protoplanetary disks have a multi-phase physical structure (see, e.g., the reviews by Woitke et al. 2009 and Dullemond \& Monnier 2010)\nocite{woitke09,dullemond10} 
seen in molecular line observations from the far-UV to the millimeter.  \citet{dutrey01} reviewed the mm-wave CO measurements of outer protoplanetary disks, but understanding the complex CO excitation structure with radius requires additional tracers of
the bulk molecular material within 100 AU of the star~\citep{greaves04}. Emission from OH, CO, CO$_{2}$, H$_{2}$O, and other
biologically important species have been used to trace warm gas ($T$~$\sim$~few~$\times$~10$^{2}$~--~2000 K) in the inner 
disk~\citep{najita03,salyk08,bethell09}.  

Metal-bearing molecules trace molecular hydrogen (H$_{2}$), the primary constituent of gas giant planets.  However, 
H$_{2}$ can be observed best in the far-UV (912~--~1650~\AA) bandpass, where the dipole-allowed molecular emission spectrum is primarily photo-excited (``pumped'') by stellar Ly$\alpha$ photons~\citep{ardila02,herczeg02}.   The Ly$\alpha$-pumping route requires that the second excited vibrational level ($v$~=~2) of H$_{2}$ have an appreciable population, which requires that the molecules reside in a warm ($T(H_{2})$~$>$~2000 K) gas layer close to the star~\citep{herczeg06}. Most likely, this photoexcited H$_{2}$ is not physically associated with the bulk of the colder H$_{2}$ that goes into planet formation.  The homonuclear nature of H$_{2}$ means that rovibrational transitions are dipole forbidden, making direct detection of the cool-H$_{2}$ component observationally challenging.  Electron-impact excitation has been suggested as a means of directly probing the H$_{2}$ in the planet-forming regions of the disk~\citep{bergin04}.  This mechanism requires stellar X-rays to create a distribution of photoelectrons that pump the molecules into 
excited electronic states.  The H$_{2}$ then relaxes, producing a characteristic far-UV cascade spectrum of discrete emission lines and quasi-continuous spectral features during the dissociation of H$_{2}$~\citep{ajello84,abgrall97}.   
\citet{ingleby09} proposed this mechanism to explain a portion of the low-resolution far-UV spectra from a sample of classical and weak-lined (low gas content) T Tauri stars
observed with the {\it Hubble Space Telescope}-ACS and -STIS.  Low-resolution and intermediate S/N in existing far-UV observations make it difficult to separate the electron-impact H$_{2}$ signal from Ly$\alpha$-pumped H$_{2}$, atomic emission, CO, and the underlying accretion continuum.  We lump these emissions together under the term ``continuum'' due to their mostly unresolved structure in previous studies.  
Fortunately, the observational picture has improved dramatically with the installation of the Cosmic Origins Spectrograph (COS) on $HST$.  $HST$-COS combines very low detector backgrounds with moderate spectral resolution ($\Delta$$v$~$\approx$~17 km s$^{-1}$).   The low background permits measurements of the true continuum shape, including an assessment of the contribution from a hot accretion continuum.  The resolution of COS also permits the identification and separation of the spectral components,  enabling a more robust comparison with models of the processes that govern the planet-forming regions of CTTS disks.

In this paper, we use COS to study in detail the far-UV continuum emission from these objects for the first time.   We focus on two objects that we propose are prototypes for electron-impact H$_{2}$ emission (V4046 Sgr) and continuum emission produced in the hot accretion shock (DF Tau).  CO $A$~--~$X$ band emission also contributes to the far-UV line and continuum spectrum in these systems. First results on CO emission will be presented in Paper II~(K. France~--~in preparation).  In \S2, we describe the COS observations and data reduction.  In \S3, we describe empirical fits to the data and compare these fits with model H$_{2}$ spectra.  \S4 describes our measurements of the collisionally excited H$_{2}$ and accretion continua.  We also discuss discrepancies between the new COS observations and predictions of our electron-impact H$_{2}$ models.  We then discuss in \S5 the relevance of these processes to the physical state of the disk and the local far-UV radiation field.  Finally, we present a brief summary of our work in \S6.  

\section{Targets and Observations}	

DF Tau and V4046 Sgr are both binary pre-main sequence systems with gas-rich disks.  DF Tau is composed of two early M stars (0.68 and 0.51~$M_{\odot}$; White \& Ghez 2001)  with an age estimated to be between $\sim$~0.1~--~2 Myr (see Ghez et al. 1997 for a discussion).\nocite{ghez97,white01}  DF Tau A is thought to be the stronger contributor to the far-UV output from the system with a mass accretion rate approximately an order of magnitude greater than DF Tau B~\citep{herczeg06}.   The DF Tau disk is observed approximately edge-on ($i$~$\sim$~80~--~85\arcdeg; Johns-Krull \& Valenti 2001; Ardila et al. 2002).\nocite{krull01,ardila02}  Emission from hot gas lines and Ly$\alpha$-pumped H$_{2}$ have been studied using existing  $HST$-STIS echelle and $FUSE$ spectra~\citep{ardila02,herczeg06}, and detailed analysis of the radiative transfer interplay between \ion{H}{1} and H$_{2}$ using COS will be presented by H. Yang (submitted).
The 0.1~--~4.5~keV luminosity of DF Tau is $\sim$~9~$\times$~10$^{29}$ erg s$^{-1}$~\citep{walter81}.  The outer disk has a dust mass~$\sim$~130 $M_{\oplus}$~\citep{andrews05} and mm-wave CO has been detected~\citep{greaves05}.
DF Tau shows evidence of a gas-rich inner disk~\citep{najita03}, 
although a search for extended near-IR inner dust disk has returned a null detection~\citep{karr10}.  

\begin{figure}
\begin{center}
\hspace{-0.25in}
\epsfig{figure=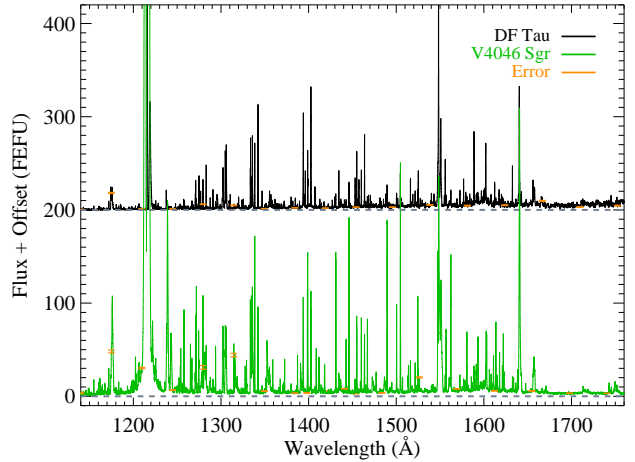,width=2.65in,angle=90}
\caption{\label{cosovly} The far-UV spectra of DF Tau ($black$) and V4046 Sgr ($green$) in the 1140~--~1760~\AA\ bandpass, obtained by combining the G130M and G160M data and smoothed to 1 spectral resolution element for display purposes.  Emission lines from hot gas (\ion{Si}{4}, \ion{C}{4}, \ion{He}{2}, etc) and Ly$\alpha$-pumped H$_{2}$ dominate the spectrum in the far-UV bandpass.  The flux is plotted in FEFU ($\equiv$~femto-erg flux unit, 1 FEFU~=~1 $\times$~10$^{-15}$ ergs cm$^{-2}$ s$^{-1}$ \AA$^{-1}$).
The DF Tau spectra have been offset by +200 FEFU for clarity.  The gray dashed line denotes the flux zero point in both observations.  
}
\end{center}
\end{figure}

\begin{deluxetable*}{cccccc}
\tabletypesize{\footnotesize}
\tablecaption{$HST$-COS observing log. \label{cos_obs}}
\tablewidth{0pt}
\tablehead{
\colhead{Object}  &   \colhead{R. A. (J2000)} &  \colhead{Dec. (J2000) }& \colhead{Date} & \colhead{COS Mode} & \colhead{T$_{exp}$ (s)} 
}
\startdata	
DF Tau & 04$^{\mathrm h}$ 27$^{\mathrm m}$ 02.81$^{\mathrm s}$ &  +25\arcdeg\ 42\arcmin\ 22.1\arcsec &   2010 January 11	 &  	G130M 	&  	 4828	 \\
DF Tau & 04$^{\mathrm h}$ 27$^{\mathrm m}$ 02.81$^{\mathrm s}$ &  +25\arcdeg\ 42\arcmin\ 22.1\arcsec &   2010 January 11	 &  	G160M 	&  	 5549      \\
V4046 Sgr & 18$^{\mathrm h}$ 14$^{\mathrm m}$ 10.49$^{\mathrm s}$ &  -32\arcdeg\ 47\arcmin\ 34.2\arcsec &   2010 April 27 	 &  	G130M 	&  	 4504	 \\
V4046 Sgr & 18$^{\mathrm h}$ 14$^{\mathrm m}$ 10.49$^{\mathrm s}$ &  -32\arcdeg\ 47\arcmin\ 34.2\arcsec &   2010 April 27	 &  	G160M 	&  	 5581      \\
 \enddata

\end{deluxetable*}

V4046 Sgr is a close binary comprised of two mid-K stars (0.91 and 0.87~$M_{\odot}$; Stempels \& Gahm 2004) with a circumbinary disk at an intermediate inclination angle ($i$~$\sim$~35\arcdeg, Quast et al. 2000; Rodriguez et al. 2010).\nocite{quast00,rodriguez10} It has an age between 4~--~12 Myr, depending on its possible membership in the $\beta$~Pic moving group~\citep{quast00,kastner08}.  
It was studied as part of the $IUE$ pre-main sequence star atlas~\citep{krull00}.  While $FUSE$ data reveal H$_{2}$ emission from the Lyman and Werner levels,  a search for mid-IR emission from cooler H$_{2}$ has returned only upper limits~\citep{carmona08}.
The 0.45~--~6.0 keV luminosity is estimated to be $\sim$~1~--~5~$\times$~10$^{30}$ erg s$^{-1}$ (G{\"u}nther et al. 2006; and \S4.3).  V4046 Sgr has a molecule-rich outer disk ($M_{gas}$~$\sim$~110~$M_{\oplus}$, $M_{dust}$~40~$\sim$~$M_{\oplus}$; Rodriguez et al. 2010), although the inner disk shows evidence of dust clearing~\citep{jensen97}. 
The lack of near-IR excess categorizes V4046 Sgr as a ``transitional'' disk, although the ongoing accretion and far-UV molecular emission suggest that a molecular gas reservoir is present in the inner disk.  

DF Tau and V4046 Sgr were observed with the medium resolution, far-UV (G130M and G160M) modes of COS on 11 January and 27 April 2010, respectively.  They were each observed for a total of four orbits, with two orbits per grating.  In order to achieve continuous spectral coverage and minimize fixed pattern noise, observations in each grating were made at four central wavelength settings ($\lambda$1291, 1300, 1309, and 1318 for G130M and $\lambda$1589, 1600, 1611, and 1623 for G160M) in the default focal plane position (FP-POS~=~3).  This combination of grating settings covers the complete 1136~$\leq$~$\lambda$~$\leq$~1796~\AA\ bandpass at a resolving power of $R$~$\approx$~18,000\footnote{The COS LSF experiences a wavelength dependent non-Gaussian form due to mid-frequency wave-front errors produced by the polishing errors on the $HST$ primary and secondary mirrors; {\tt http://www.stsci.edu/hst/cos/documents/isrs/}}.  Near-UV imaging target acquisitions were performed 
through the COS primary science aperture using MIRRORB.
Table 1 provides a log of the COS observations acquired as part of this study. Table 2 summarizes relevant system parameters, including accretion diagnostics measured in this work.  The data have been reprocessed with the COS calibration pipeline, CALCOS\footnote{We refer the reader to the cycle 18 COS Instrument Handbook for more details: {\tt http://www.stsci.edu/hst/cos/documents/handbooks/current/cos\_cover.html}} v2.12, and combined with the custom IDL coaddition procedure described by~\citet{danforth10}.  Figure 1 displays the combined 1140~--~1760~\AA\ spectra of both systems.  The flux units used on this and subsequent figures are FEFUs\footnote{see \S1.1.2 of the Cycle 17 COS Instrument Handbook} (1 FEFU~=~1 $\times$~10$^{-15}$ ergs cm$^{-2}$ s$^{-1}$ \AA$^{-1}$) .   We use multi-epoch archival optical data spectra as supporting observations of the spectral region around the Balmer break.  

\begin{deluxetable*}{lcccccccc}
\tabletypesize{\footnotesize}
\tablecaption{System Parameters. \label{obj_params}}
\tablewidth{0pt}
\tablehead{ \colhead{Object} & \colhead{Spectral Type} & \colhead{Age} & \colhead{Inclination} & \colhead{log$_{10}$ ($L_{X}$)\tablenotemark{a}} & \colhead{Ref.} 
 & \colhead{log$_{10}$ ($G_{cont}$/$G_{o}$)\tablenotemark{b}} & 
\colhead{log$_{10}$($L_{CIV}$)} & \colhead{log$_{10}$($\dot{M}$$_{acc}$)} \\
\colhead{} & \colhead{} & \colhead{(Myr)} & \colhead{ } & \colhead{(erg s$^{-1}$)} & \colhead{} & \colhead{($a$~=~1 AU)} & \colhead{(erg s$^{-1}$)} & \colhead{($M_{\odot}$ yr$^{-1}$)} 
}
\startdata 
DF Tau & M0.5 + M3 & 0.1~--~2 Myr & $>$~80\arcdeg & 29.9 & {\it 1,2,3,4} & 6.8 & 
30.26 & -7.1\tablenotemark{c} \\

V4046 Sgr & K5 + K5 & 4~--~12 Myr & 35\arcdeg & 30.0~--~30.7 & {\it 5,6,7,8} & 5.6 & 
29.25 & -7.9\tablenotemark{c} \\
\enddata
\tablenotetext{a}{For DF Tau, $L_{X}$ is the total 0.1~--~4.5 keV luminosity~\citep{walter81}.  For V4046 Sgr, $L_{X}$ is estimated for the total 0.45~--~6.0 keV luminosity~(see \S4.3). }
\tablenotetext{b}{$G_{o}$ is the strength of the average interstellar radiation field, evaluated over the 912~--~2000~\AA\ bandpass ($G_{o}$~=~1.6~$\times$~10$^{-3}$ erg cm$^{-2}$ s$^{-1}$; Habing 1968).\nocite{habing68} } 
\tablenotetext{c}{Determined from the empirical relationship between C IV emission and the mass accretion rate, Eqn. (2) of~\citet{krull00}. \\
References: (1)~\citet{ardila02}, (2)~\citet{krull01}, (3)~\citet{ghez97}, 
(4)~\citet{white01}, (5)~\citet{quast00}, (6)~\citet{stempels04}, (7)~\citet{kastner08}, 
(8)~\citet{rodriguez10} } 
\end{deluxetable*}

\section{Analysis} 

\subsection{Emission from H$_{2}$ and the Inner Accretion Disk}

The strongest features in the far-UV spectra of DF Tau and V4046 Sgr are the  emission lines from hot gas and photoexcited H$_{2}$ (Figure 1).  The hot gas lines (e.g., \ion{N}{5}, \ion{Si}{4}, \ion{C}{4}, etc) are produced in the magnetically active atmospheres~\citep{bouvier07} of the central stars and at (or near) the shock interface where the accretion stream impacts the stellar surface~\citep{gunther08}.  The strong, discrete Lyman and Werner band H$_{2}$ lines are mostly fluoresced by H$_{2}$ pumping transitions that coincide with the broad Ly$\alpha$ profiles in the these systems, although coincidences with other stellar lines also contribute~\citep{wilkinson02,france07b}.  We see that the hot gas and photoexcited H$_{2}$ lines are superimposed upon a mostly continuous, faint underlying emission spectra.  \citet{ingleby09} argued that this emission is a combination of accretion luminosity from near the hot accretion spots and emission from electron-impact excited H$_{2}$.   
The electron-impact H$_{2}$ can be further decomposed into discrete bound-bound emission and quasi-continuous emission from dissociative transitions to the ground electronic state, which is characteristic of the electron impact process for electron distributions with energies $\gtrsim$~15 eV~\citep{ajello91}.  
In both systems, there is also a contribution from CO emission which will be addressed in Paper II. 
In Figure 2, we show a breakdown of these components for V4046 Sgr.  The accretion continuum is represented as the blue dashed line, and a model H$_{2}$ spectrum is shown superimposed as a dotted red line.  Because the vast number (several hundreds) of photoexcited H$_{2}$ lines complicate the comparison of the broad band spectral characteristics of the data with the model, we isolate spectral regions that are free of known photoexcited H$_{2}$ lines.  These ``continuum bins'' have widths of $\approx$~0.6~--~1~\AA, and are shown as purple diamonds in Figure 2.   

\begin{figure}
\begin{center}
\hspace{-0.25in}
\epsfig{figure=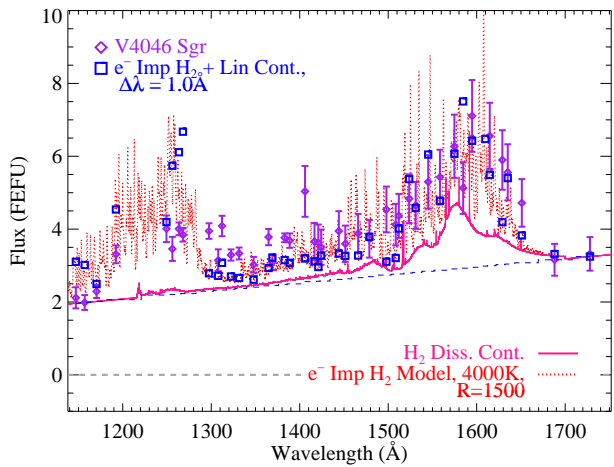,width=2.65in,angle=90}
\caption{\label{cosovly} The binned continuum spectrum of V4046 Sgr ({\it purple diamonds  with error bars}) is compared with a synthetic electron-impact H$_{2}$ spectrum ($blue$~$squares$) created for parameters from the literature [$N$(H$_{2}$)~=~10$^{19}$ cm$^{-2}$, $T$(H$_{2}$)~=~4000K, $E_{e}$~=~20eV].  The model has been offset by an underlying linear continuum, shown as the blue dashed line.  The decomposition of the model into discrete and quasi-continuous emission is also shown ({\it red dotted and pink solid lines, respectively}).  The discrete line emission has been shown at a decreased resolution for display purposes.  We conclude that the observed continuum spectrum can be decomposed into an accretion continuum (dashed blue line) and emission from H$_{2}$.  We note that the model overpredicts the observed emission at $\lambda$~$\lesssim$~1300~\AA, as discussed in \S3.3.  
}
\end{center}
\end{figure}

The ``continuum'' spectra of the two systems are qualitatively different.  Figure 3 ($left$) shows the continuum spectrum of DF Tau, corrected for reddening ($A_{V}$~=~0.6; Herczeg \& Hillenbrand 2008) using the interstellar curve of~\citet{ccm}.  The far-UV data show a mostly linear rise with wavelength that we attribute to  accretion emission. 
We show a linear fit to the data as the dashed red line and several optical spectra of DF Tau, representative of the high (purple), intermediate (blue), and low (green) states of the object.  We include a scaled version of the intermediate spectrum as the dashed black line to illustrate that the linear fit to the far-UV data could account for the observed continuous spectrum of DF Tau below the Balmer break (see \S4.2 for a discussion of the spectral shape of this emission).  We attribute the dispersion of the far-UV points at $\lambda$~$>$~1500~\AA\ to contamination by unresolved H$_{2}$ lines (both photo- and electron-excited), CO band emission, and the lower S/N data at the red end of the far-UV bandpass where the COS effective area decreases.  The V4046 Sgr spectrum includes emission with a shape qualitatively similar to that expected for electron-impact excited H$_{2}$ (see Figure 3, $right$). As we illustrated in Figure 2, this emission is superimposed upon an underlying linear continuum.
We do not make an extinction correction for V4046 Sgr because $A_{V}$ is negligible~\citep{stempels04}.  
We consider DF Tau as an example of a CTTS with hot accretion emission and V4046 Sgr as representative of a CTTS with electron-excited H$_{2}$.  
In the following subsections, we describe the modeling procedure we used to develop synthetic H$_{2}$ spectra for a quantitative analysis  of the electron-impact spectrum.  The characteristics of each process are considered in more detail in \S4.  

\subsection{Synthetic Electron-Impact H$_{2}$ Spectra}

We created synthetic electron-impact H$_{2}$ spectra for comparison with the COS observations following the formalism described by~\citet{liu96} and~\citet{wolven97}.  Our spectral synthesis code includes emission from photoexcited H$_{2}$~\citep{france10a}, but we have decoupled the various emission components in order to study separately the discrete and quasi-continuous emission spectrum from collisionally excited H$_{2}$. We describe the calculation of electron-impact spectra here.   The model includes emission from the $B$~--~, $B^{'}$~--~, $B^{''}$~--~, $C$~--~, $D$~--~, and $D^{'}$~--~$X$
electronic transitions, including Lyman band populations coming from $E,F$~--~$B$ cascades.  
We explicitly include transitions to predissociating states and vibrational states that 
result in dissociation ($v^{''}~>$~14, the vibrational continuum) and independently save the dissociation spectra. Transition probabilities and wavelengths for the $B$,  $B^{'}$, $C$, and $D$ were taken from the literature~\citep{abgrall93a,abgrall93b,abgrall94}, 
the  $B^{''}$ and  $D^{'}$ parameters were calculated by combining the branching ratios of the  $B^{'}$ and $D$ states with the appropriate energy level spacings~\citep{kwok85,glass84,huber79}. We computed transition probabilities for the $E$ and $F$ cascades  using the Franck-Condon factors presented by~\citet{lin74}.  
The emission rate from level of $nv'J'$ to level $v''J''$ of the ground electronic state is given by
\begin{equation}
S(n'v'J' \rightarrow v''J'')~=~ R(n'v'J')
[1 - \eta(n'v'J')]\frac{A_{n'v'J' \rightarrow v''J''}}{A_{n'v'J'}}	,
\end{equation}
where $R(n'v'J')$ is the excitation rate into the excited electronic level, 
$\eta(n'v'J')$ is the efficiency for predissociation in the excited electronic state, and the ratio of transition probabilities 
($A_{nvJ}$) is the branching ratio~\citep{liu96}.~\nocite{abgrall93a,abgrall93b}

\begin{figure*}
\begin{center}
\hspace{+0.0in}
\epsfig{figure=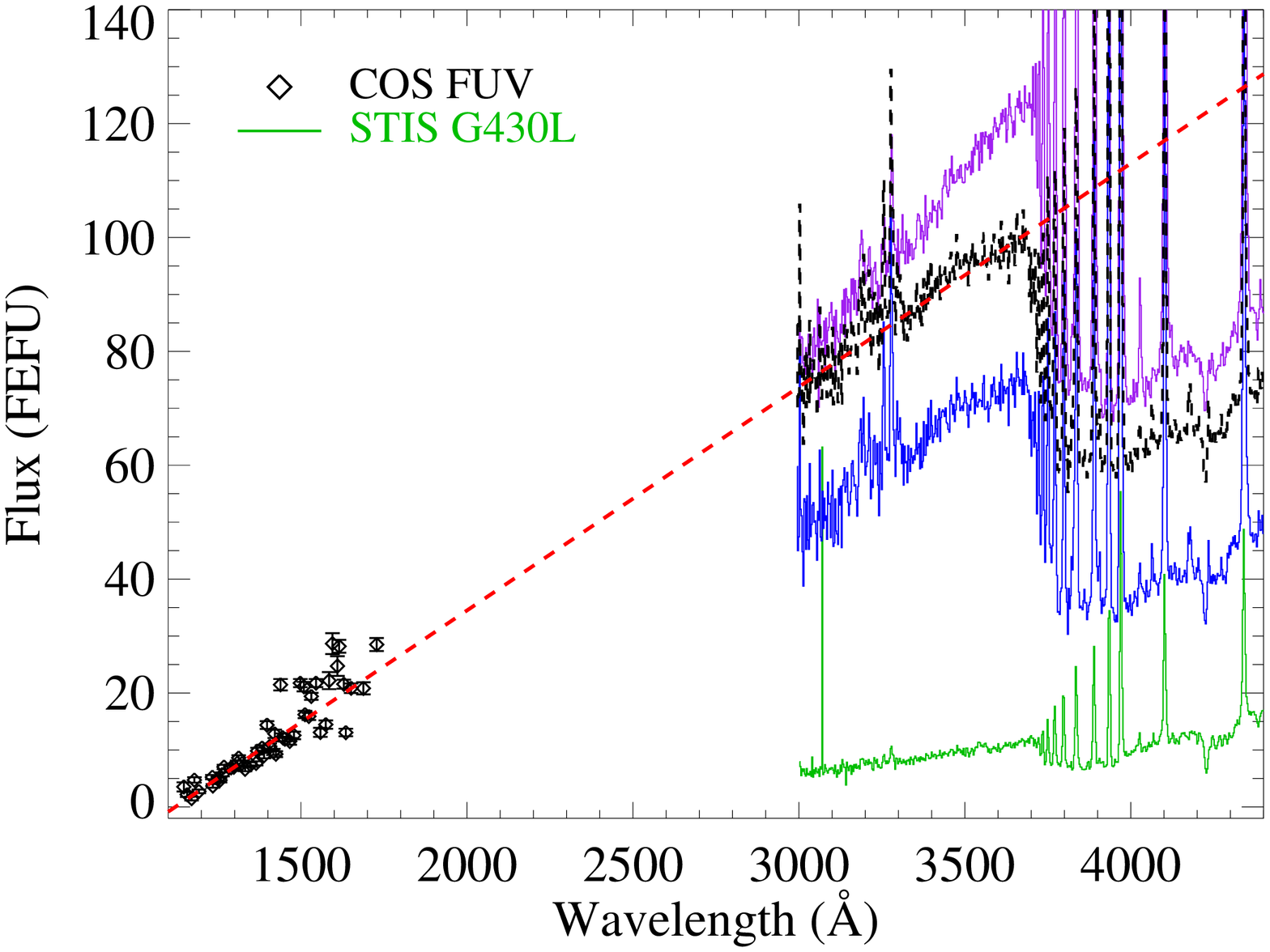,width=2.35in,angle=90}
\epsfig{figure=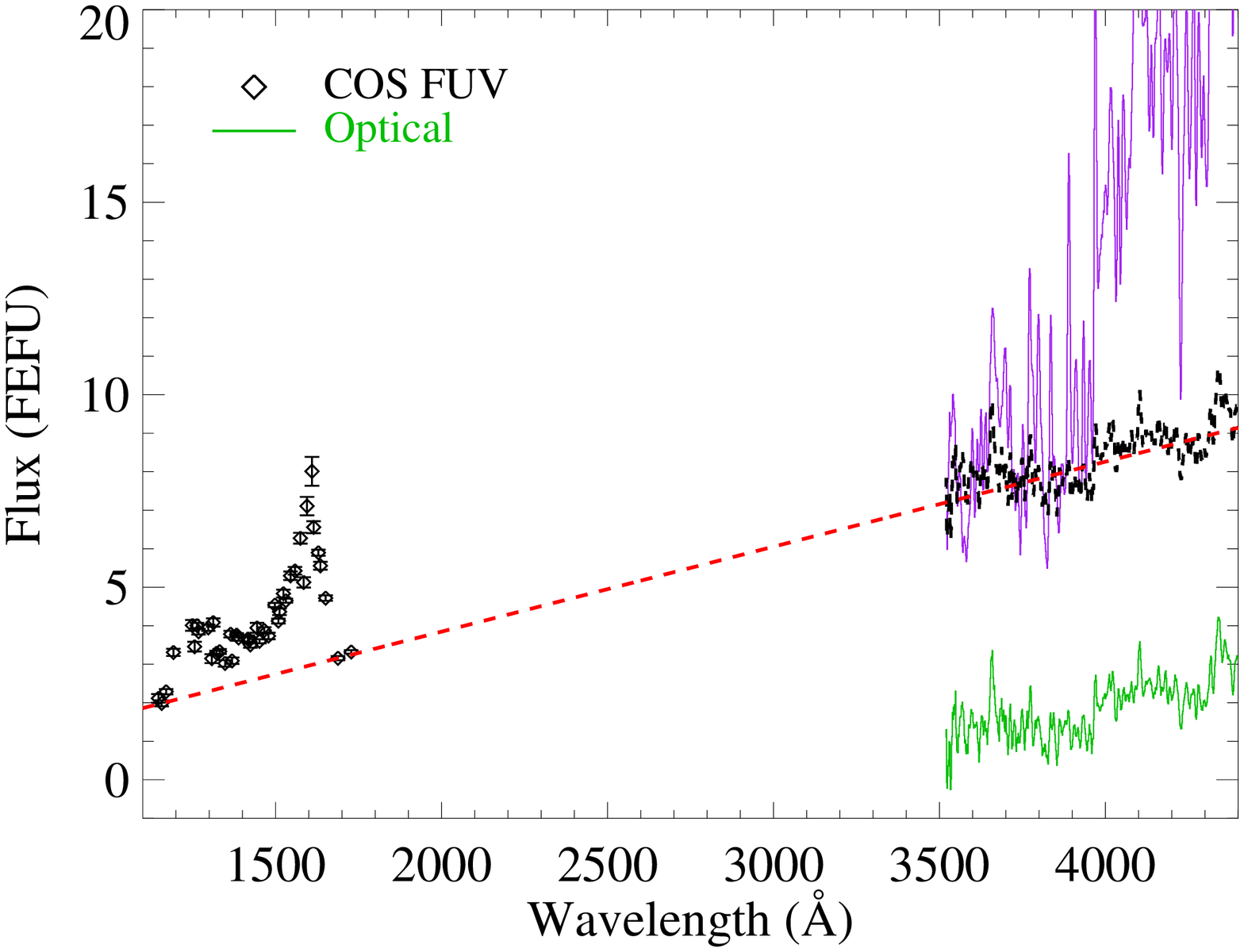,width=2.35in,angle=90}
\caption{\label{cosovly} 
The far-UV and near-UV/optical accretion continuum in DF Tau ($left$) and V4046 Sgr ($right$).
 A linear fit to the far-UV spectra ({\it red dashed line}) is extrapolated to longer wavelengths, and compared with the ``high'', ``intermediate'', and ``low''  optical emission states (plotted in $purple$, $blue$, and $green$, respectively).  
The optical spectrum displayed as the black dashed line has been offset to show that a linear continuum can describe the accretion emission from the far-UV to the Balmer break. The extra emission above the dashed red line in the DF Tau data near 1600~\AA\ is most likely contamination from photo-excited CO and 
H$_{2}$ emission, while for V4046 Sgr the extra emission above the linear fit is dominated by emission from electron-impact H$_{2}$.  
}
\end{center}
\end{figure*}

The excitation rate, $R$, is proportional to the column density in the ground electronic state, $N_{vJ}$, the cross-section connecting 
the ground state to the rovibrational level of the excited electronic state, $\sigma_{vJ \rightarrow n'v'J'}$, and the electron flux, $F_{e}$~\citep{liu98}, 
\begin{equation}
 R(n'v'J')~=~F_{e}N_{vJ}\sigma_{vJ \rightarrow n'v'J'}.
\end{equation}
Given the uncertainties about the source of the free electrons and their geometry with respect to the spatial distribution of the H$_{2}$, we do not attempt to reproduce the absolute flux of the electron-impact emission in this work. Thus the electron flux can be assumed to be a scale factor.  

The cross-sections can be measured experimentally, but as we are not solving for the absolute flux of the electron-impact emission lines in the spectra of V4046 Sgr, we follow the procedure described by~\citet{liu98} and calculate relative excitation rates described by the equation 
\begin{equation}
 \sigma_{vJ \rightarrow n'v'J'}~=~\frac{\Omega_{vJ \rightarrow n'v'J'}}{\omega_{vJ}E_{e}},
\end{equation}
where $E_{e}$ is the electron energy, $\omega_{vJ}$ is the degeneracy of the ground electronic state, and $\Omega_{vJ \rightarrow nv'J'}$ is the collision strength given by the analytic expression presented in Eqn (4a) of~\citet{liu98}.   

For simplicity, we assume a single population of collisionally excited gas.  We input the ground state thermal temperature [$T(H_{2})$], the column density [$N(H_{2})$], and the electron energy.  We created a grid of models in [$T(H_{2})$,$N(H_{2})$,$E_{e}$]-space, including ten temperatures between 300~--~5000 K, five column densities in the range 10$^{16}$~--~10$^{20}$ cm$^{-2}$, and ten electron distributions in the range 14~--~5000 eV.   The lower limit at $E_{e}$~$\geq$~14eV is set by the minimum energy needed to populate the $C$, $E$,  and $F$ levels in our model.  Numeric errors occur when these states are not appreciably populated.  Physically, the absolute cross-sections for electron excitation into the Lyman and Werner bands are very low at $E_{e}$~$<$ 14eV, with $\sigma_{H2}$~$\lesssim$~3~$\times$~10$^{-18}$ cm$^{2}$ at $E_{e}$~$\lesssim$~12eV  (compared to $\sigma_{H2}$~$>$~2~$\times$~10$^{-16}$ cm$^{2}$ at 20~$\lesssim$~$E_{e}$~$\lesssim$~200eV; Liu et al. 1998).~\nocite{liu98}  

\begin{figure}
\begin{center}
\hspace{-0.25in}
\epsfig{figure=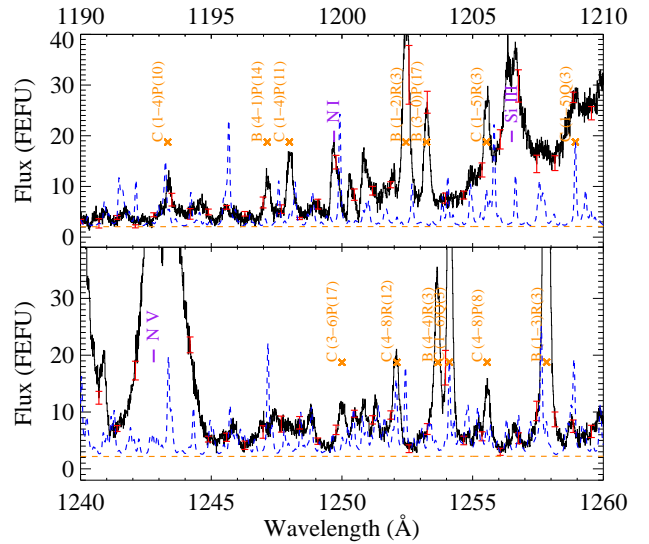,width=2.65in,angle=90}
\vspace{+0.5in}
\caption{\label{cosovly} A comparison of the full-resolution short-wavelength observations of V4046 Sgr ($black$) with the predicted discrete line emission from an unattenuated electron-impact H$_{2}$ model ({\it blue dashed line}; see \S3.3).  Despite the agreement at $\lambda$~$>$~1400~\AA, many of the predicted features are not observed.  Atomic species (e.g., \ion{N}{1}, \ion{Si}{3}, and \ion{N}{5}) are labeled in purple, and emissions from photoexcited H$_{2}$ are labeled in orange.
The photoexcited lines are listed in Table 2.   
}
\end{center}
\end{figure}

\begin{deluxetable*}{lcccc}
\tabletypesize{\footnotesize}
\tablecaption{V4046 Sgr photoexcited H$_{2}$ emission lines in Figures 4 and 6. \label{lya_lines}}
\tablewidth{0pt}
\tablehead{
\colhead{Line ID\tablenotemark{a}} & \colhead{$\lambda_{obs}$} & \colhead{$\lambda_{rest}$}  & 
\colhead{Pumping Transition\tablenotemark{b}}  &  \colhead{$\lambda_{pump}$}    \\ 
    & (\AA) & (\AA) &  &  (\AA)   }
\startdata
$C$ (1~--~4) P(10)	& 	1193.37 	&   1193.34   & 	(1~--~5) R(8) 	&   1214.62	 \\
$B$ (4~--~1) P(14)	& 	1197.15 	&   1197.14   & 	(4~--~2) R(12) 	&   1213.68	 \\
$C$ (1~--~4) P(11)	& 	1198.00 	&   1197.99   & 	(1~--~5) R(9) 	&   1217.00	 \\
$B$ (1~--~2) R(3) 	& 	1202.44 	&   1202.45   & 	(1~--~2) P(5) 	&   1216.07	 \\
$B$ (3~--~0) P(17)	& 	1203.24 	&   1203.24   & 	(3~--~1) R(15) 	&   1214.47	 \\
$C$ (1~--~5) R(3)		& 	1205.53 	&   1205.53   & 	(1~--~5) P(5) 	&   1216.99	 \\
$C$ (1~--~5) Q(3)	& 	1208.90 	&   1208.93   & 	(1~--~1) Q(3)\tablenotemark{c} 	&   1031.86	 \\
\tableline
$C$ (3~--~6) P(17)	& 	1249.99 	&   1250.00   & 	(3~--~5) P(17) 	&   1216.74	 \\
$C$ (4~--~8) R(12)	& 	1252.05 	&   1252.10   & 	(4~--~7) R(12)\tablenotemark{d} 	&   1220.47	 \\
$B$ (4~--~4) R(3)	 	& 	1253.63 	&   1253.67   & 	(4~--~3) P(5) 	&   1214.78	 \\
$C$ (1~--~6) Q(3)	& 	1254.11 	&   1254.11   & 	(1~--~1) Q(3)\tablenotemark{c} 	&   1031.86 \\
$C$ (4~--~8) P(8)		& 	1255.55 	&   1255.56   & 	(4~--~7) P(8) 	&   1219.95	 \\
$B$ (1~--~3) R(3)	 	& 	1257.81 	&   1257.83   & 	(1~--~2) P(5) 	&   1216.07	 \\

 \enddata

	\tablenotetext{a}{$B$ denotes a Lyman band ($B$$^{1}\Sigma^{+}_{u}$~--~$X$$^{1}\Sigma^{+}_{g}$) transition, $C$ denotes a Werner band ($C$$^{1}\Pi_{u}$~--~$X$$^{1}\Sigma^{+}_{g}$) transition.} 
	\tablenotetext{b}{H$_{2}$ lines excited by the broad stellar \ion{H}{1} Ly$\alpha$ emission line, unless otherwise noted. } 
	\tablenotetext{c}{Pumped through a coincidence with the stellar \ion{O}{6} $\lambda$1032~\AA\ resonance line. } 
	\tablenotetext{d}{The observed line may contain additional H$_{2}$ fluorescence excited through coincidence with \ion{H}{1} Ly$\beta$ and/or
\ion{Si}{3} $\lambda$1206.50~\AA. } 
\end{deluxetable*}

\begin{figure}
\begin{center}
\hspace{-0.25in}
\epsfig{figure=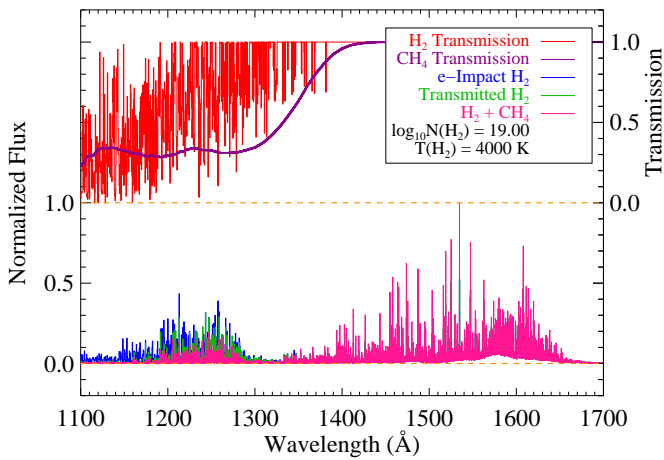,width=2.65in,angle=90}
\caption{\label{cosovly} ($top$) Transmission functions of methane and optically thick H$_{2}$ through the disk of V4046 Sgr. 
($bottom$) The synthetic electron-impact H$_{2}$ spectrum is plotted in blue, the synthetic spectrum attenuated by H$_{2}$ self-absorption is shown in green, and the model attenuated by H$_{2}$~+~CH$_{4}$ is shown in magenta.  
The blue, green, and magenta curves are the same at $\lambda$~$\gtrsim$~1400~\AA, showing that molecular absorption most strongly impacts the observations in the G130M bandpass. 
}
\end{center}
\end{figure}

\subsection{Comparing the H$_{2}$ Models with the V4046 Sgr Spectra}

The model spectra computed on a 0.01~\AA\ grid are convolved with the appropriate COS linespread function (LSF) for comparison with our spectroscopic observations.  
The electron-impact spectrum of H$_{2}$ consists of both discrete and quasi-continuous emission features.  The continuum features are strongest at $\lambda$~$>$~1350~\AA~\citep{abgrall97}, while the electron-impact spectral energy distribution is mostly discrete at the blue end of the COS bandpass ($\lambda$~$<$~1300~\AA).   Figure 2 shows the qualitative agreement of the V4046 Sgr data with the H$_{2}$ model, and we anticipate that a broad-band $\chi^{2}$ minimization approach can be used to estimate the H$_{2}$ parameters of the system.  

In order to explore what additional processes should be included in the models prior to fitting the data, we compare the broad spectral colors
with the intrinsic color ratio ($CR$) used to characterize electron penetration into gas giant atmospheres, $CR$~$\equiv$~$I(1550-1620$~\AA)/$I(1230-1300$~\AA)~$\sim$~1.1~\citep{yung82,gustin06}, where 
\begin{equation}
 I(\Delta\lambda)~=~\int^{\lambda_{2}}_{\lambda_{1}} F_{H2} d \lambda
\end{equation}
and $F_{H2}$ is the electron-impact H$_{2}$ flux.  Confusion from the overlap with hundreds of strong photoexcited H$_{2}$ emission lines in the CTTS spectra forced us to evaluate the data in $\approx$~1~\AA\ bins, as described in \S3.1.  Using this binning, we find an intrinsic model color ratio of $CR_{mod}$~$\sim$~1.3\footnote{However, when our models are evaluated over the entire ``blue'' and ``red'' bandpasses for reasonable input parameters, we find that the total integrated flux is actually slightly greater is the sub-1400~\AA\ band  [$I(1140-1400$~\AA)~$>$~$I(1400-1680$~\AA)], which is a consequence of the large contribution from discrete features.  
 This qualitative result is true for our model spectra over a wide range of physical parameters, that is, the ``blue'' and ``red'' fluxes are approximately equal. }.
The observed color ratio in V4046 Sgr is $CR_{obs}$~=~2.1~$\pm$~0.2, indicating that the observations are significantly redder than their intrinsic colors.  
Jovian color ratios in this range are typically interpreted as internal reddening  in the upper atmosphere~\citep{dols00,gustin06}.  

The model predicts the spectrum of the discrete lines that have been observed in short-wavelength far-UV spectra of the gas giant planets~\citep{gustin04,gustin09}.   As a first-order check on the discrete lines, we compare in Figure 4 a section of the full-resolution COS G130M data with an electron-impact H$_{2}$ model.  The model, overplotted as the dashed blue line, assumes parameter values typical of previous electron-impact studies of CTTS disks [$N$(H$_{2}$)~=~10$^{19}$ cm$^{-2}$, $T$(H$_{2}$)~=~4000K, $E_{e}$~=~20eV], although this electron distribution has a slightly higher energy than those favored by~\citet{ingleby09}.  We discuss the electron energy distribution in \S4.3.    Figure 4 shows that many discrete features predicted by the model are not seen in the data. 
Because of this disagreement and the reddening of the broad-band colors, we conclude that calculations of electron-impact spectra taken from the literature are alone inadequate for direct comparison with the COS observations. In the following subsections, we therefore develop a more detailed physical picture that produces agreement with both constraints provided by  the observed H$_{2}$ spectrum.   
Table 3 presents a list of the photoexcited H$_{2}$ lines identified in Figure 4.

\subsubsection{H$_{2}$ Self-Absorption}

A strong motivation for studying the electron-impact excited H$_{2}$ spectrum in protoplanetary disks is that the more easily observable Ly$\alpha$-pumped H$_{2}$ emission is likely produced on the surface of the disk~\citep{herczeg06}, whereas the X-rays needed to produce the non-thermal electron population can penetrate deeper into the planet forming region~\citep{bergin04,ingleby09}.  Thus, the electron-impact excited gas may be a useful probe of the molecules that are in the actively planet-forming regions of the disk.   
The role of radiative transfer in determining the resulting UV spectrum measured at Earth has been considered for radiative excitation of H$_{2}$ in the disk surface~\citep{herczeg04} but not for the collisionally excited H$_{2}$ in the inner disk.  We first consider reddening by dust. We discount strong internal reddening because grains in the disk have likely grown appreciably during the  few Myr age typical of CTTS disks~\citep{furlan05,apai05}, and there is evidence that large grains produce a relatively ``gray'' extinction across the far-UV bandpass~\citep{fm88,ccm}.  Grain growth and settling has also been shown to significantly reduce the overall opacity to UV photons in disks older than $\sim$~1 Myr~\citep{vasyunin10}.  
Also, if these inner disk grains play a large role, they must also extinguish stellar emission in edge-on systems, which is clearly not the case (e.g., the strong continuum and hot gas lines are observed in DF Tau).

\begin{figure*}
\begin{center}
\hspace{+0.0in}
\epsfig{figure=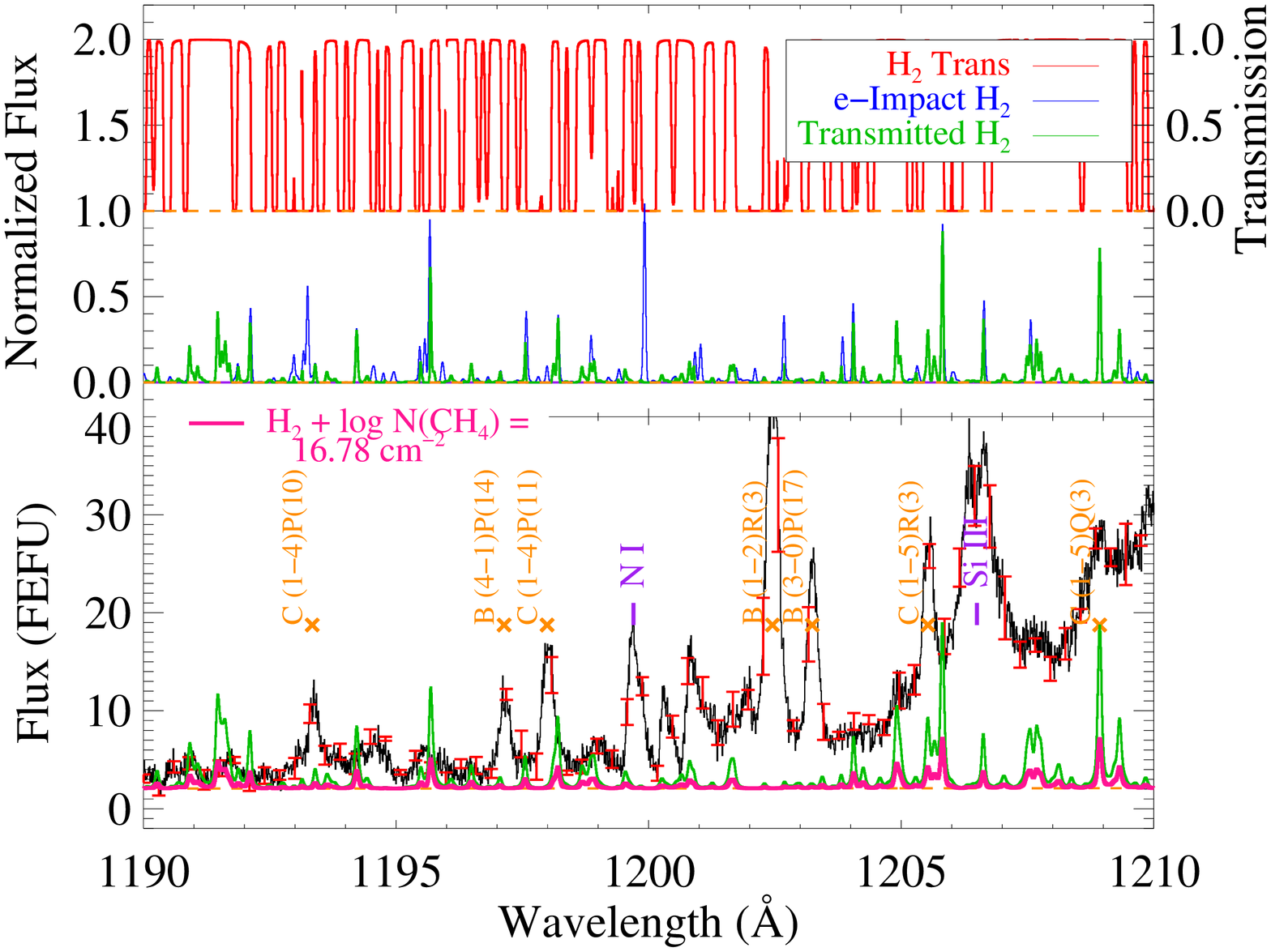,width=3.75in,angle=90}
\epsfig{figure=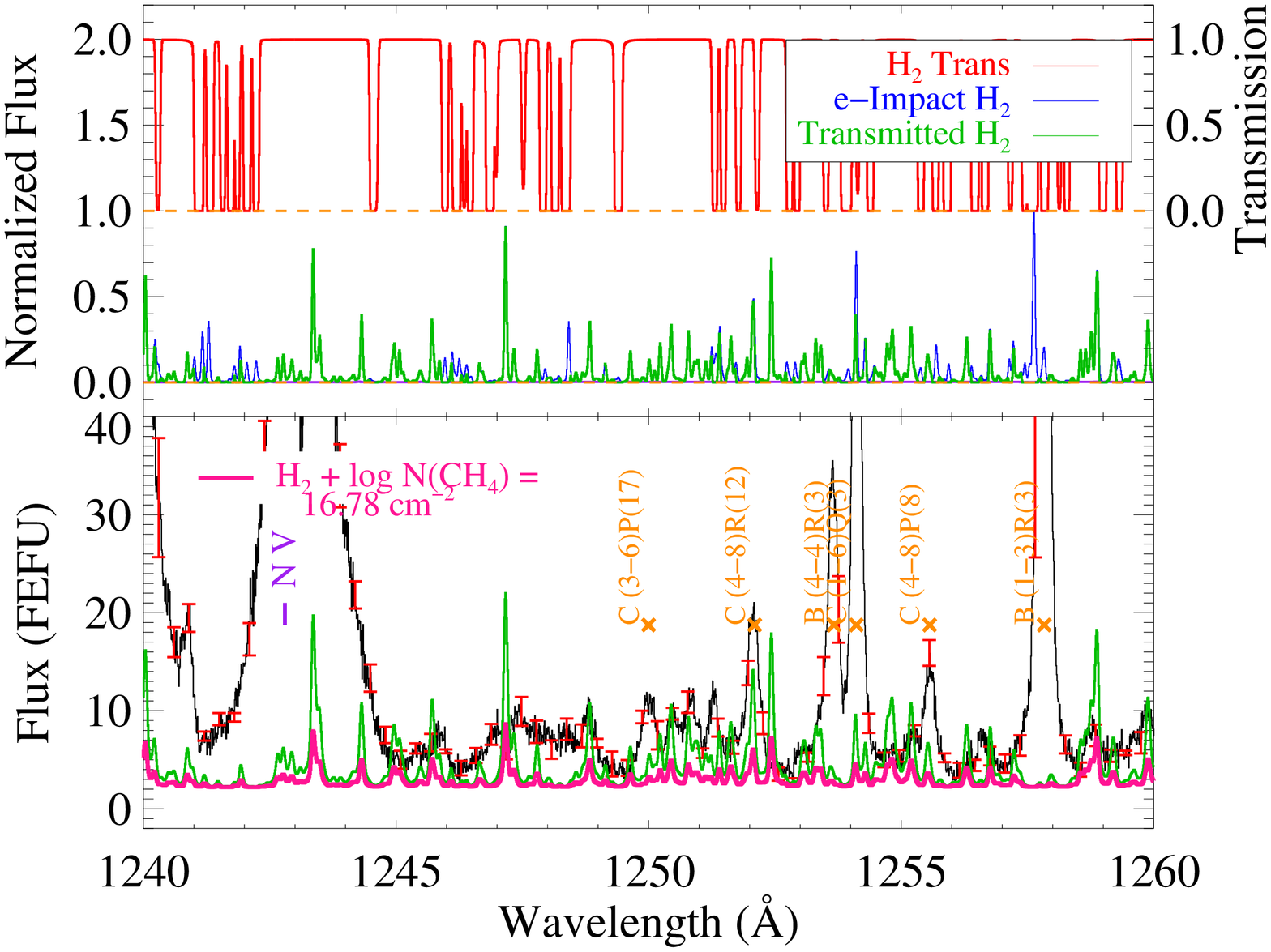,width=3.75in,angle=90}
\vspace{+0.25in}
\caption{\label{cosovly} A combination of Figures 4 and 5 illustrating the effects of H$_{2}$ and CH$_{4}$ absorption at full spectral resolution.  The V4046 Sgr spectra are plotted in black.  
The red H$_{2}$ transmission curves show that even at high temperatures and relatively high column densities [$T$(H$_{2}$)~=~4000 K, $N$(H$_{2}$)~=~10$^{19}$ cm$^{-2}$] there are significant windows through which H$_{2}$ emission can escape in the COS G130M bandpass ({\it shown in green}).  By contrast, more continuous absorption by hydrocarbons can effectively remove the short wavelength flux from H$_{2}$ ({\it shown in magenta}) while allowing the electron-impact excited H$_{2}$ emission observed in the COS G160M bandpass to escape (see Figure 5).  
Assuming that all of the hydrocarbon absorption can be attributed to CH$_{4}$, we find that $N$(CH$_{4}$)~$\geq$~6~$\times$~10$^{16}$ cm$^{-2}$ is required to suppress the discrete line emission to the observed level.  
}
\end{center}
\end{figure*}

An effect that must be considered is self-absorption by H$_{2}$ in the outer layers of the disk.  At first consideration, it seems plausible that H$_{2}$ photons emitted in the disk plane are re-absorbed before they can escape.  This would naturally explain both the red $CR$ and the absence of discrete emission features because self-absorption would remove individual lines from the resultant emission spectrum, thereby reddening the model spectrum.   As cool H$_{2}$ [$T$(H$_{2}$)~$\lesssim$~700 K] has appreciable optical depth only in the first few rotational levels ($J$~$\lesssim$~6), this gas will not produce significant absorption out of excited vibrational levels.  This cool H$_{2}$ resembles the well-studied spectrum of H$_{2}$ in interstellar clouds (e.g., Rachford et al. 2002; Burgh et al. 2007)\nocite{rachford02,burgh07} and cannot provide appreciable opacity in the COS bandpass ($\lambda$~$\gtrsim$~1150~\AA).  A population of hot H$_{2}$ [$T$(H$_{2}$)~$>$~2000 K] would be required. We have tested this hypothesis by creating a tunable absorbing ``screen'' of H$_{2}$ that will attenuate the radiation produced by the electron-impact model.  We created H$_{2}$ absorption spectra for 2000~$\leq$~$T$(H$_{2}$)~$\leq$~5000 K by calculating the Boltzmann distribution for the ground electronic state and used the H$_{2}$ optical depth templates from~\citet{mccandliss03} to compute transmission functions as a function of temperature and column density.  We included all rovibrational states in the range (0~$\leq$~$J$~$\leq$~14) and  (0~$\leq$~$v$~$\leq$~3).  Even larger temperatures would be required to thermally populate higher states, but we do not consider such temperatures because they are significantly above the $\approx$~4500 K H$_{2}$ dissociation temperature~\citep{lepp83}. Figure 5 shows the transmission function of H$_{2}$ in red at the top, while the intrinsic ($blue$) and transmitted ($green$) H$_{2}$ emission spectra are shown at the bottom.

In general, we find that H$_{2}$ self-absorption is insufficient to account for the color and discrete feature discrepancies described in \S3.3. 
Our results for H$_{2}$ self-absorption of discrete lines are shown as illustrative examples in Figure 6 for typical temperatures [$T$(H$_{2}$)~=~4000K] and column densities [$N$(H$_{2}$)~=~10$^{19}$ cm$^{-2}$] obtained by~\citet{ingleby09}.   The top panels of Figure 6 show the computed H$_{2}$ transmission spectrum in red, the unattenuated electron-impact model in blue, and the transmitted H$_{2}$ spectrum in green.  
We find that while several of the transitions are indeed completely absorbed, many of the strongest transitions are transmitted. It is these emission lines that we would expect to observe.  The lower panels of Figure 6 show that these emissions are not observed, 
even though the COS sensitivity curve peaks in the $\lambda$~$\approx$~1250~\AA\ window displayed in Figures 4 and 6.  The primary reason that self-absorption alone cannot explain the discrepancy is that the H$_{2}$ population created by electron-impact is highly non-thermal, whereas upper layers of the disk are mostly described by a warm thermal distribution (2000~$\lesssim$~$T$(H$_{2}$)~$\lesssim$~3000 K; Herczeg et al. 2006) with deviations caused by UV photoexcitation.\nocite{herczeg06}  We conclude that while H$_{2}$ self-absorption must be included, it is not the dominant factor in determining the shape of the emitted electron-impact H$_{2}$ spectrum in V4046 Sgr.

\subsubsection{Hydrocarbon Absorption}

In the previous section, we discussed the contribution of dust and H$_{2}$ self-absorption to the spectral characteristics of the observed spectrum of V4046 Sgr; in particular, that the long-wavelength ($\lambda$~$\gtrsim$~1450~\AA) continuum could be described by emission from H$_{2}$ excited by non-thermal electrons, but there is a lack of discrete emission lines at shorter wavelengths.  We have determined that neither dust nor H$_{2}$ self-absorption can attenuate synthetic electron-impact H$_{2}$ spectra to the extent observed.  

We now consider absorption by more complex molecules, CO, H$_{2}$O, and hydrocarbons.   We first assume that the electron-excited H$_{2}$ is located in the disk and that the H$_{2}$ emission must pass through a ``screen'' of these more complex molecules to reach an external observer. 
We require that the attenuating screen removes the short-wavelength H$_{2}$ emission, but not the flux at longer wavelengths.  
While CO is observed in UV spectra of CTTS disks (e.g., Paper II), the $A$~--~$X$ spectrum that is the most prominent CO feature at far-UV wavelengths has a characteristic band structure.  Even for very high temperatures, the interband spectral regions have low opacity.  As a simple quantitative test for the importance of CO absorption within the disk, we created a template CO optical depth spectrum for column density $N$(CO)~=~10$^{17}$ cm$^{-2}$ at $T$~=~1000~K, with spectral binning of 0.05~\AA.  Even for this warm, high column density medium, we find that only 
approximately 7\% of the spectral bins have $\tau_{CO}$~$>$~0.1.  Thus, there are ample low-opacity windows through which the underlying H$_{2}$ spectrum can escape.  A similar argument can be made for the spectrum of water.  Laboratory studies have identified individual vibrational modes of the H$_{2}$O Rydberg series in the 6~--~11 eV energy region~\citep{mota05}.  The far-UV cross-sectional spectrum is characterized by strong bands in the 9.9~--~10.8~eV (1250~$\gtrsim$~$\lambda$~$\gtrsim$~1150~\AA) region where we require substantial attenuation of the H$_{2}$ electron-impact spectrum.  However, because the interband spectral regions have cross-sections of $\sigma_{H2O}$~$\lesssim$~2~$\times$~10$^{-18}$ cm$^{-2}$, very large column densities of water would be required for attenuation of the electron-impact H$_{2}$ emission which is thought to be produced in the inner disk~\citep{bergin04}.   
The required column densities needed to reach $\tau_{H2O}$~$\approx$~1 are larger than those found in observational and model studies of H$_{2}$O in the mid-IR spectra of protoplanetary disk systems~\citep{carr08,bethell09}.  Finally, H$_{2}$O has a prominent absorption band in the 7~--~8~eV range~\citep{diercksen82,mota05}, which is coincident with the broad 1575, 1608~\AA\ spectral features characteristic of electron-impact excitation of H$_{2}$.  This dissociative H$_{2}$O absorption feature has absolute cross-sections that are comparable to or stronger than those in the short-wavelength windows described above.  Thus water column densities high enough to attenuate the bluest wavelengths in our data would also eliminate the flux at 1600~\AA, which is not observed.   

We suggest that a more likely scenario involves the absorption of H$_{2}$ emission by hydrocarbons in the planet-forming disk.  This scenario is analogous to what has long been observed in far-UV auroral spectra of gas giant planets: a mostly electron-impact H$_{2}$ spectrum that is ``reddened'' by hydrocarbon absorption, mostly from methane, ethane, and acetylene (CH$_{4}$, C$_{2}$H$_{6}$, and C$_{2}$H$_{2}$; Clarke et al. 1980; Yung et al. 1982).~\nocite{clarke80,yung82}  In particular, CH$_{4}$ and C$_{2}$H$_{6}$ have absorption cross-sections that meet the required criteria:  ($i$) strong ($\sigma$~$>$ 10$^{-17}$ cm$^{-2}$), quasi-continuous absorption below $\lambda$~=~1400~\AA, and  ($ii$) negligible effect on the output spectrum  ($\sigma$~$<$ 10$^{-18}$ cm$^{-2}$) at the long-wavelength end of the COS bandpass.   
In the Jovian atmosphere, methane is often assumed to be the most important species regulating the broad-band far-UV spectral signature of H$_{2}$~\citep{kim97,grodent01}.  We compute the total methane opacity required to match the observed flux limits from the non-detection of discrete H$_{2}$ electron-impact features in the COS data at $\lambda$~$<$~1300~\AA.  We first scaled the 1575~--~1608~\AA\ model emission spectrum to match the observed flux.  We then included self-absorption following the procedure described above, and then applied a methane transmission function to the data.  
We find that $N$(CH$_{4}$)~$\gtrsim$~6~$\times$~10$^{16}$ cm$^{-2}$ is sufficient to attenuate the short-wavelength discrete features within the errors of the observed V4046 Sgr data.  The relative abundances of the hydrocarbons in the disk are unknown, and a similar exercise assuming a purely ethane screen requires 
$N$(C$_{2}$H$_{6}$)~$\gtrsim$~4~$\times$~10$^{16}$ cm$^{-2}$.  The difference is due to the somewhat larger cross-sections of ethane in the 1140~--~1300~\AA\ bandpass. 


We caution the reader that our assumption of pure methane absorption in the disk was chosen by analogy to planetary atmospheres.  Although protoplanetary disks are the building blocks of these atmospheres, the analogy is uncertain.  We do not have a strong observational constraint on the choice of methane over ethane, although the former is almost always found to be the dominant source of far-UV attenuation in the Jovian atmosphere~\citep{dols00}.  A combination of hydrocarbons is undoubtedly a more likely model, and there may also be additional species that we have not considered in the present analysis.  Whatever the exact makeup of the disk, some sort of absorbing screen is required if the usual assumption that H$_{2}$ electron-impact emission is a combination of bound-bound and dissociative transitions is valid in these disks.  Modeling the far-UV radiative transfer in these disks is required to obtain a more quantitative understanding of the spectra we now routinely observe with $HST$-COS.    

\section{Discussion}

\subsection{Electron-Impact H$_{2}$ Emission in V4046 Sgr}

In \S3.3, we described how the $\lambda$~$<$~1400~\AA\ discrete emission lines from the electron-impact H$_{2}$ excitation cascade are absorbed and/or scattered out of the line-of-sight.  In order to measure the properties of the collisionally excited H$_{2}$ spectrum, we performed a $\chi^{2}$ minimization analysis to compare the binned V4046 Sgr data at $\lambda$~$>$~1400~\AA\ with the grid of electron-impact H$_{2}$ models described in \S3.2.  The free parameters are   
[$N$(H$_{2}$), $T$(H$_{2}$), $E_{e}$] and a linear fit to the underlying continuum, which we will discuss in the following subsection.  We find that the electron-impact excited H$_{2}$ has a temperature of $T(H_{2})$~=~3000$^{+1000}_{-500}$ K, a column density of $N$(H$_{2}$)~$\sim$~10$^{18}$ cm$^{-2}$, an electron energy of $E_{e}$~$\sim$~50~--~100~eV, but the only firm constraint is on the temperature. By restricting the fits to $\lambda$~$>$~1400~\AA, we avoid the 1200~--~1300~\AA\ region where our models predict a dependence on $N$(H$_{2}$).   We find a minimum near $E_{e}$~$\sim$~50~--~100 eV, but higher energies are generally consistent with the data for  $N$(H$_{2}$)~=~10$^{16-19}$ cm$^{-2}$.  In Figure 7, we show representative fits to the  $\lambda$~$>$~1400~\AA\ data.  We find that 
the best fit parameters not only display the lowest reduced $\chi^{2}$, but they also show reasonable fits to the eye.  For comparison, we also display a low energy case ($E_{e}$~=~14 eV). 

\begin{figure}
\begin{center}
\hspace{-0.25in}
\epsfig{figure=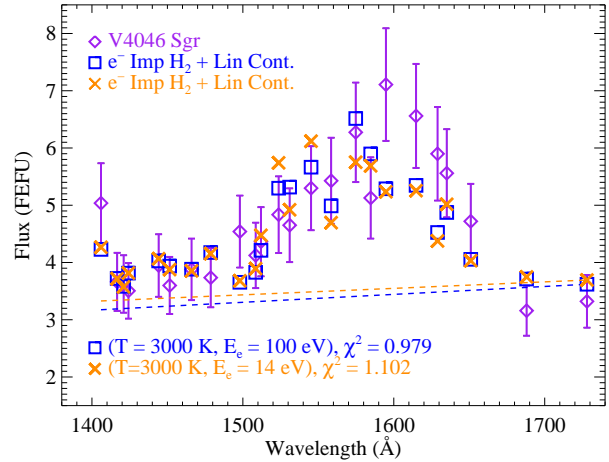,width=2.65in,angle=90}
\caption{\label{cosovly} The best-fit model of the electron-impact excited H$_{2}$ emission in the spectrum of V4046 Sgr.  Due to the attenuation of the short-wavelength discrete emission, we restrict the fits to $\lambda$~$>$~1400~\AA.  The best fit model parameters are $T(H_{2})$~=~3000$^{+1000}_{-500}$ K, N($H_{2}$) $\sim$~10$^{18}$ cm$^{-2}$, and electron energies in the range $E_{e}$~$\sim$~50~--~100 eV.  The H$_{2}$ column density is only weakly constrained by this analysis.  We also show the low energy model ($E_{e}$~=~14 eV) for the best fit temperature and column density.  The underprediction of the model flux near 1600~\AA\ is due to not including CO emission from the disk.}
\end{center}
\end{figure}

A major complication to all of these fits is the spectral overlap between the electron-impact emission spectrum of H$_{2}$ and CO $A$~--~$X$ band emission.  With our broad binned approach, we can avoid strong discrete lines from the stellar atmosphere and accretion shock, as well as photoexcited H$_{2}$.  The CO emission is distributed in wide vibrational band structures with non-zero intraband flux. Using the binned continuum bands described above, we necessarily include CO emission.  We will present models of this emission in Paper II.  For now, we note that the strongest contaminant to the pure H$_{2}$ electron-impact spectrum is the CO (0~--~1) $\lambda$1597 \AA\ band that fills in the spectral valley between the H$_{2}$ emission peaks at 1575 and 1608~\AA.  This excess emission tends to push the fits to higher temperatures, where the dissociation continuum broadens, or to very low electron energies ($E_{e}$~$<$~14 eV), where dissociation is less important.  This may explain some of the discrepancies between our results and the best-fit parameters [$T(H_{2})$~$\sim$~5000 K; $E_{e}$~$\sim$~12 eV] presented by~\citet{ingleby09}.

\subsection{Accretion Continuum}

Figure 3 shows the spectra of our targets from the far-UV to the optical.  The black dotted optical spectrum has been scaled from an archival observation to show that the linear fit to the far-UV data alone (dashed red line) can plausibly also describe the shape of the $\lambda$~$<$~3600~\AA\ Balmer continuum used to measure the accretion rates in low-mass pre-main sequence stars with disks~\citep{calvet98,herczeg09}. 
We fit the far-UV spectra with both a power-law and a straight line.  For both spectra a simple linear fit of the form $F_{\lambda}$~=~$m$$\lambda$~+~$b$ produced the best match to the data.  We find [$m$,$b$] = [0.0393, -44.0250] and [0.0026, 0.0918] for DF Tau and V4046 Sgr, respectively, where the units of $m$ and $b$ are FEFU \AA$^{-1}$ and FEFU, where 1 FEFU~=~1 $\times$~10$^{-15}$ erg cm$^{-2}$ s$^{-1}$ \AA$^{-1}$.  The linear slope for DF Tau is $\sim$~15 times that found for V4046 Sgr.  While DF Tau is known to have a much larger accretion rate (2~$\lesssim$~$\dot{M}$$_{acc}$~$\lesssim$~1300~$\times$~10$^{-9}$ $M_{\odot}$ yr$^{-1}$; Hartigan et al. 1995; Herczeg et al. 2006; Herczeg \& Hillenbrand 2008)\nocite{hartigan95,herczeg06,herczeg08} than V4046 Sgr ($\dot{M}$$_{acc}$~$\sim$~3~$\times$~10$^{-11}$ $M_{\odot}$ yr$^{-1}$; G{\"u}nther and  Schmitt 2007)\nocite{gunther07} a direct comparison between accretion rates from the literature and the far-UV accretion continuum slopes is unreliable given the different measurement methods and temporal variability.  

We can estimate the mass accretion rates directly from the COS observations using empirical scaling relations between the \ion{C}{4} luminosity ($L_{CIV}$) and $\dot{M}$$_{acc}$.   
The largest systematic source of uncertainty to this approach is the relative contribution from the stellar transition region emission to the observed \ion{C}{4} profile.  We use equation 2 from~\citet{krull00} to estimate the mass accretion rate in these systems at the time of our continuum observations, assuming that the \ion{C}{4} surface flux in excess of a saturated stellar component ($F_{CIV}$~$>$~10$^{6}$ erg cm$^{-2}$ s$^{-1}$) is produced at the accretion shock.   

We fit the \ion{C}{4} profiles with a multi-component line-profile employing the appropriate COS LSF.  Assuming distances of 140 and 70 pc for DF Tau and V4046 Sgr, respectively, we find total $L_{CIV}$~=~1.81~$\pm$~0.09~$\times$~10$^{30}$ erg s$^{-1}$ and 1.76~$\pm$~0.04~$\times$~10$^{29}$ erg s$^{-1}$. 
The error bars are the measurement errors, which are very small for these bright lines. These are the time-averaged luminosities over the total G160M observing time. 
The \ion{C}{4} fluxes were found to be slowly varying over the $\approx$ 3 hours of G160M exposures.  We measure \ion{C}{4} count rate changes of +7.0 and -13.6\% for DF Tau and V4046 Sgr from the COS microchannel plate two-dimensional spectrograms, evaluated in 50s time intervals.  

These \ion{C}{4} luminosities can be explained by mass accretion rates of 7.8~$\times$~10$^{-8}$ $M_{\odot}$ yr$^{-1}$ and 1.3~$\times$~10$^{-8}$ $M_{\odot}$ yr$^{-1}$ for DF Tau and V4046 Sgr, respectively.  While our accretion rate for DF Tau is only slightly larger than the range quoted by~\citet{herczeg08}, our measurement for V4046 Sgr is considerably larger than the value quoted by~\citet{gunther07}.  The large discrepancy with the~\citet{gunther07} accretion rate may be attributable to a larger contribution from transition region \ion{C}{4} in V4046 Sgr, however it seems unlikely that values of the mass accretion rate as low as $\dot{M}$$_{acc}$~$\sim$~3~$\times$~10$^{-11}$ $M_{\odot}$ yr$^{-1}$ are consistent with our observations of V4046. 
This is not entirely surprising as CTTS mass accretion rates based on 
X-ray observations typically differ from optical/UV determinations by more than an order of magnitude due to absorption in the high-density regions where shock X-rays are produced (see e.g., Sacco et al. 2010 and references therein).\nocite{sacco10}
The relevant accretion measurements are given in Table 2.  


We suggest that future observations for which far-UV, near-UV, and optical observations are obtained in close time proximity are required to measure accurately both the far-UV continuum and the accretion rate from the Balmer continuum.  These observations will tie together the observed far-UV  accretion continuum in these objects with the well-calibrated accretion rate diagnostic.

\subsection{Outstanding Issues Regarding Electron-Impact Excitation of H$_{2}$}	

There is some uncertainty in the assumption that electron-excited H$_{2}$ is 
the dominant molecular continuum emission mechanism for V4046 Sgr and other young circumstellar disks.  Our models and analysis support a picture in which intermediate energy electrons excite H$_{2}$ in the inner disk, which then radiates and is observed superimposed upon an underlying continuum from the accretion shock. Since X-rays are invoked to create the photoelectrons~\citep{draine78} needed to collisionally excite the H$_{2}$, we must ask if X-rays are sufficient for this mechanism to operate.  We can measure the total flux from electron-excited H$_{2}$ in V4046 Sgr, $F_{H2}$~$\sim$ 8~$\pm$~4~$\times$~10$^{-13}$ erg cm$^{-2}$ s$^{-1}$ (approximately 2~$\times$ the integrated 1400~--~1660~\AA\ H$_{2}$ flux shown in Figure 7 in order to account for the implied $\lambda$~$<$~1400~\AA\ emission, \S3.3).  If we assume an inner disk electron-impact H$_{2}$ emitting region (0.2~$\leq$~$a_{H2}$~$\leq$~0.5 AU; see \S5.2 for additional discussion), the collisionally excited H$_{2}$ luminosity is 
$L^{cont}_{H2}$~$\sim$~9~$\pm$~5~$\times$~10$^{28}$ erg s$^{-1}$, or   
$\sim$~6~$\pm$~3~$\times$~10$^{39}$ photons s$^{-1}$ for a fiducial H$_{2}$ wavelength of 1400~\AA.  

High resolution $Chandra$ spectra of V4046 Sgr show that the accretion shock flux is approximately half of the observed value in TW Hya~\citep{gunther06,gudel07}.  
Scaling from the observed 0.45~--~6.0 keV X-ray luminosity ($L_{X}$) of TW Hya~\citep{kastner02}, we estimate the total X-ray luminosity for V4046 Sgr is $L_{X}$~$\sim$~1~--~5~$\times$~10$^{30}$ erg s$^{-1}$, or 0.6~--~3~$\times$~10$^{39}$ photons s$^{-1}$
for a fiducial X-ray energy of 1 keV.  1 keV is the approximate peak energy of the V4046 Sgr X-ray spectral energy distribution~\citep{gunther07}.  The grain photoelectric yields for the sub-micron sized grains that are present at the inner edge of transitional dust disks~\citep{eisner06} are of order unity for photons with $E$~$\gtrsim$~1~keV~\citep{weingartner06}.  Thus, we find that V4046 Sgr produces sufficient X-ray luminosity to create enough photoelectrons to excite the observed H$_{2}$ emission from the inner disk. 

While this electron-impact H$_{2}$ interpretation is energetically plausible for the observed emission,  it may not be adequate to describe the full far-UV emission spectrum of V4046 Sgr.  Absorption by a hydrocarbon screen appears to reproduce both the observed absorption of the short-wavelength discrete lines and the unobscured H$_{2}$ emission at $\lambda$~$\gtrsim$~1400~\AA, but verification of this result requires additional study.  A more speculative scenario involves a very high column density, high temperature self-absorbing screen of H$_{2}$ that could produce nearly continuous absorption in the COS G130M band while not absorbing the long wavelength emission arising from dissociative transitions.  Another possible scenario is a ``non-traditional'' electron-impact H$_{2}$ emission process in which dissociative yields of order unity are achieved.  This process would produce the quasi-continuous long wavelength emission without most of the energy being distributed among the discrete states of the Lyman and Werner bands.  As the canonical H$_{2}$ dissociation fraction from the Lyman and Werner bands is $\sim$~0.15~\citep{stecher67,shull82,abgrall97}, large dissociative yields would require substantially fine-tuned molecular and electron distributions.  
It seems unlikely that either of these scenarios are physically possible.  
The other prominent long-wavelength emission source in these disks is CO, although CO alone probably cannot account for the observed far-UV molecular spectra of CTTS disks.


We point out that a spectrograph such as COS is essential for a comprehensive analysis of such systems.  The low background and high-sensitivity are both important because the continua in these systems have characteristic flux levels of a few~$\times$~10$^{-15}$ erg cm$^{-2}$ s$^{-1}$ \AA$^{-1}$ or less.  Moderate spectral resolution is also required for the detection of individual discrete lines predicted by electron-impact models.  Moreover, when protoplanetary systems are observed $without$ heavy internal reddening, spectral coverage at 
$\lambda$~$<$~1300~\AA\ is important because this region probes the discrete line emission from the Werner bands of H$_{2}$ whose spectrum is a more sensitive 
diagnostic of the electron energy distribution.  Without this wavelength region, spectral fits to the data 
will not account for the Werner lines, biasing the results toward very low energy electron distributions.  As the threshold energies for the Lyman and Werner bands are 11.37 and 12.41 eV~\citep{ajello84}, respectively, Lyman band continuum emission in the absence of discrete Werner band lines requires a highly-tuned electron distribution.  Because Ly$\alpha$ production from the dissociation of H$_{2}$ drops sharply at 14.7eV~\citep{ajello91}, discrete lines should be even more important relative to the continuum at very low electron energies.

\citet{herczeg06} presented high-quality $HST$-STIS E140M observations of DF Tau. While the spectral resolution of these data is sufficient to analyze the brighter Ly$\alpha$ pumped H$_{2}$ emission, the accretion and H$_{2}$ continua are too confused by the instrumental noise floor to make a clear separation.  Conversely, \citet{ingleby09} presented a large sample of CTTS targets using the ACS-SBC, but the low spectral resolution makes the detection of discrete molecular features impossible.  Neither of these instruments has sufficient throughput in the $\lambda$~$<$~1300~\AA\ band to measure the Werner emission lines of H$_{2}$.

\section{Probing the Far-UV Radiation Environment and Molecular Gas of Inner CTTS Disks}

\subsection{The 912~--~2000~\AA\ Radiation Field and Molecular Dissociation}

The high sensitivity and spectral resolution of COS allows us to isolate the accretion shock continuum in pre-main sequence circumstellar disk systems.  This higher temperature continuum radiation field is similar to that discussed for longer wavelengths by~\citet{calvet98} and the multi-component fit to the UV spectrum of TW Hya by~\citet{costa00}.  In Figure 8, we show the continuum spectrum of DF Tau, whose central object is an M2 binary system (see \S2), with a photospheric temperature of $T_{eff}$~$\sim$~3500 K~\citep{hartigan95,krull00}.  The effective temperature of the accretion emission in DF Tau is ~$\approx$~10$^{4}$ K, which creates a local radiation field in the planet-forming region of the disk with a shape similar to that of a Herbig Ae star.  To illustrate this point, we compare the scaled $HST$-STIS spectrum of AB Aur in Figure 8 ($green$) with the binned DF Tau continuum spectrum.  AB Aur and DF Tau have similar ages. AB Aur is known to be surrounded by a molecule-rich disk (H$_{2}$ and CO; Roberge et al. 2001), and there is evidence that planets are forming in the system~\citep{oppenheimer08}.\nocite{roberge01}  A critical difference between DF Tau and AB Aur (spectral type A0Ve, $T_{eff}$~$\approx$~10$^{4}$ K) is the temperature of the UV radiation field.  

\begin{figure}
\begin{center}
\hspace{-0.5in}
\epsfig{figure=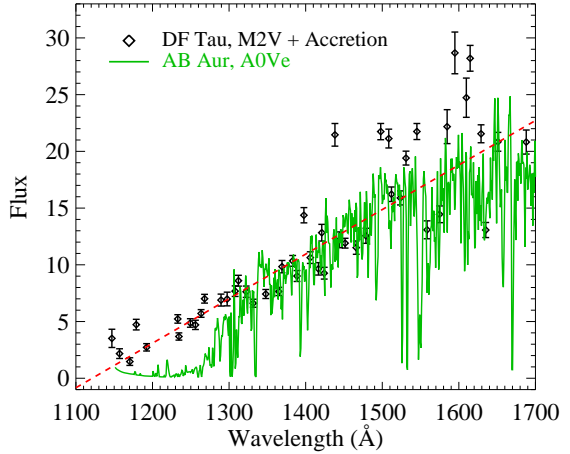,width=2.65in,angle=90}
\caption{\label{cosovly} Comparison of the spectral shape of the far-UV accretion shock emission in DF Tau with photospheric emission from the Herbig Ae star AB Aur.  The linear fit to the DF Tau continuum is shown as the red dashed line.  The strong continuum emission from DF Tau is an effective radiation field that is $\sim$~3 times hotter than the radiation from the primary star, creating a source of far-UV continuum photons in the planet forming region of the disk. The enhanced flux (above the linear fit) in DF Tau near 1600~\AA\ is most likely CO emission.}
\end{center}
\end{figure}

The accretion radiation detected down to $\lambda$~=~1147~\AA\ with high confidence provides strong evidence for continuum emission at wavelengths down to the Lyman limit\footnote{The fit to the DF Tau data reaches zero at 1122~\AA, but this is clearly an artificial cut-off imposed by the simple linear model.  The data show appreciable continuum flux above the linear fit at $\lambda$~$<$~1200~\AA. }.
The shape and intensity of this far-UV radiation field have a significant impact on the evolution of a protoplanetary disk~\citep{vand08}.  In particular, dissociation of the most abundant species is controlled by photons in this critical 912~--~1150~\AA\ wavelength region.  H$_{2}$ is photodissociated through a two-step process~\citep{stecher67} starting with absorption into the Lyman bands at $\lambda$~$\leq$~1108~\AA~\citep{carruthers70}, extending to longer wavelengths for rovibrationally excited molecules.  The first strongly predissociating CO absorption system is the $E$~--~$X$ band at~$\lambda$~$\leq$~1076~\AA~\citep{vand88}, with shorter wavelength far-UV transitions dominating the dissociation as radiative transfer effects become important deeper into the disk.  The dissociation of H$_{2}$ and CO molecules produce species that can catalyze the chemistry throughout the inner disk, including neutral H, vibrationally excited H$_{2}$ (H$_{2}^{*}$), neutral and ionized C, and O.  This suggests that the boundary regions where the far-UV continuum penetrates the higher density molecular disk could be considered photodissociation regions (PDRs).
Far-UV continuum emission will contribute to the dissociation of water (as discussed in \S3.3.2), which is relevant for both giant planet formation and the terrestrial planet formation stages that follow.  Further out in the disk, this emission will control the photodissociation rate of C$_{2}$H and H$_{2}$CO, which modify the molecular abundances at larger radii~\citep{zadelhoff03}.

Far-UV continuum emission has been considered before in low-mass protoplanetary disk models, but some models assume that the $\lambda$~$<$~1400~\AA\ flux is dominated by the interstellar radiation field (ISRF) and stellar emission lines. We present, however, evidence that the absolute flux of the far-UV continuum is several orders of magnitude greater than the ISFR.  
Integrating the continuum fits (shown in Figure 3) over the far-UV bandpass (912~--~2000~\AA), we find that the absolute strength of the far-UV continua are 10$^{5-7}$ times the interstellar value ($G_{o}$~$\equiv$~1.6~$\times$~10$^{-3}$ erg cm$^{-2}$ s$^{-1}$; Habing 1968).~\nocite{habing68}  Assuming distances of 140 and 70 pc for DF Tau and V4046 Sgr, we find log$_{10}$($G/G_{o}$)~=~6.8 and log$_{10}$($G/G_{o}$)~=~5.6, respectively, at 1 AU.  We note that this is only 912~--~2000~\AA\ accretion continuum radiation field strength, not including the flux from atomic emission lines.  Our radiation field estimates are consistent with the CTTS UV intensities calculated for 100 AU by~\citet{bergin04}.  

One caveat is that this A-star-like accretion emission may not be present at all times, or at least not at the flux levels we observe.  Figure 3 shows that optical activity may be a good proxy for the level of far-UV continuum. Thus the far-UV accretion emission is most likely variable, while the external field provided by the ISRF is constant.   

\subsection{Physical Conditions of the Molecular Gas in the Inner Disk}

We find that the electron-impact excited H$_{2}$ population in V4046 Sgr is characterized by $N$(H$_{2}$)~$\sim$~10$^{18}$ cm$^{-2}$ and $T$(H$_{2}$) $\sim$~3000 K.  This temperature is high but is not unreasonable for the inner region of an X-ray irradiated disk~\citep{meijerink08}.  These authors find that this temperature can be maintained in the inner ~$\approx$~20 AU of the disk, consistent with H$_{2}$ emission excited by photoelectrons produced by the X-rays~\citep{bergin04}.  While the outer radius of the collisionally excited H$_{2}$ may be as large as 20~AU, we suggest that this process $could$ operate much closer to the central object, most likely~$\lesssim$ 1 AU~\citep{ingleby09}.  The average width of the photo-excited H$_{2}$ lines listed in Table 3 is 52~$\pm$~8 km s$^{-1}$, with evidence of double peaked structure in the highest S/N lines.  While some of this width may be due to turbulent motions within the disk, this suggests a concentration of photoexcited H$_{2}$ at 0.19~AU, for $i$~=~35\arcdeg\ and the simple assumption of a Keplerian disk where $r_{H2}$~=~($GM$/$<v_{FWHM}>$)$^{1/2}$.  

If this is the inner edge of the warm molecular disk, it is co-spatial with the inner edge of the warm dust disk (0.18~AU; Jensen \& Mathieu 1997), similar to the photo-excited H$_{2}$ distribution around the accreting brown dwarf 2M1207~\citep{france10b}.\nocite{jensen97}
While we cannot determine the spatial origin of the electron-impact excited H$_{2}$ emission conclusively, we favor an origin in the $\sim$~0.2~--~2~AU inner disk.  
While the electron-impact H$_{2}$ emission may be present at the disk midplane where planets are actively forming, the observed H$_{2}$ emission most likely originates 
between the disk midplane and the the disk surface.  
This radiation is being emitted from planet forming radii in the disk, but the collisionally excited H$_{2}$ is not tracing the cold ($T$~$\sim$~30 K), high column ($N_{H}$~$>$~10$^{22}$ cm$^{-2}$), low electron fraction ($x_{e}$~$<$~10$^{-6}$) disk midplane.   

The far-UV continuum and the ``PDR-like'' environment of the inner disk call into question the assumption that non-thermal electrons are the only  excitation source for the broad-band H$_{2}$ emission that we observe in V4046 Sgr. We note that the best fit column density and temperature values for the collisionally excited H$_{2}$ are similar to those found for the Ly$\alpha$-pumped H$_{2}$ in other CTTS disks~[$T$(H$_{2}$)~$\sim$~2500 K, $N$(H$_{2}$)~$\sim$~10$^{18-19}$ cm$^{-2}$] (Ardila et al. 2002; Herczeg et al. 2004), even though these states are not fully thermalized.\nocite{ardila02,herczeg04}  At moderate spectra resolution, the spectrum of H$_{2}$ excited by the far-UV continuum of a hot star is qualitatively very similar to the far-UV electron-impact spectrum (compare, e.g., Witt et al. 1989 and France et al. 2005 with Gustin et al. 2006).\nocite{witt89,france05a}  

Finally, in \S3.3.2 we suggested that hydrocarbons are responsible for the selective reddening of the observed H$_{2}$ electron-impact spectrum.  Methane, the most important opacity source for electron-impact H$_{2}$ emission in giant planet atmospheres, is difficult to observe at radio wavelengths owing to its lack of permanent dipole moment. However, methane has been observed in higher mass protostellar disks at abundances $\approx$~10$^{-6}$ that of hydrogen~\citep{boogert04}.  Furthermore, \citet{carr08} presented infrared spectra of acetylene emission in the disk of the CTTS AA Tau.  They posited that the large observed abundance of C$_{2}$H$_{2}$ is a sign of vertical mixing within the disk, bringing the hydrocarbons up from the midplane.  An analogous scenario fits our methane assumption. If we assume that the CH$_{4}$ is $\sim$~10$^{-6}$ times as abundant as hydrogen in the planet forming region ($N_{H}$~$\sim$~10$^{22-23}$ cm$^{-2}$) and that this gas can be transported vertically towards the disk surface, then $N$(CH$_{4}$) is sufficient to attenuate the discrete H$_{2}$ line emission in V4046 Sgr [$N$(CH$_{4}$)~$\geq$~6~$\times$~10$^{16}$ cm$^{-2}$].  Thus, we may have found indirect evidence for a reservoir of hydrocarbons in the inner disk, where they are expected to become important components of extrasolar giant planet atmospheres~\citep{sudarsky03}.

\section{Summary}
We have presented here the highest-quality observations of the far-UV continuum in CTTSs acquired to date.  Because the gas-rich CTTS phase is roughly 
co-temporal with the epoch of giant planet formation, our observations represent an important portion of the energetic radiation environment 
present during the formation of exoplanetary systems.   We characterize the spectra of hot accretion emission and electron-excited H$_{2}$ in two prototypical CTTSs.  We find that the far-UV accretion continuum can be described by a simple linear fit, which connects to the optical Balmer continuum given assumptions about the optical activity level at the time of the far-UV observations.  Accretion introduces an additional source of far-UV continuum emission that is 5~--~7 orders of magnitude stronger than the ISRF and has an effective temperature of $\sim$~10$^{4}$ K in DF Tau.  We also find that the electron-impact H$_{2}$ spectrum in V4046 Sgr is characterized by $T$(H$_{2}$)~=~3000 K and $N$(H$_{2}$)~$\sim$~10$^{18}$ cm$^{-2}$, with $E_{e}$~$\sim$~50~--~100 eV.
Finally, we propose that hydrocarbon absorption is the most likely explanation for the differential reddening inferred from the attenuation of collisionally produced H$_{2}$ emission lines below 1400~\AA.   

\acknowledgments
We appreciate helpful discussions with Greg Herczeg, Alex Brown, Laura Ingleby, and Nuria Calvet.  K. F. thanks Brian Wolven for continued guidance with the spectral synthesis code and Jacques Gustin for making his hydrocarbon cross-section database available to us. We thank Fred Walter for providing archival optical spectra of V4046 Sgr.  
This work has made use of Tom Ayres' StarCAT database, hosted at MAST.  
This work was support by NASA grants NNX08AC146 and NAS5-98043 to the University of Colorado at Boulder.





\end{document}